\documentclass[useAMS,usenatbib]{mn2e}
\usepackage{times}
\usepackage{graphicx}
\usepackage[figuresright]{rotating}

\newif\ifAMStwofonts
\AMStwofontstrue

\newcommand{\simlt}{\lower.5ex\hbox{$\; \buildrel < \over \sim \;$}}
\newcommand{\simgt}{\lower.5ex\hbox{$\; \buildrel > \over \sim \;$}}
\newcommand{\be}{\begin{equation}}
\newcommand{\ba}{\begin{eqnarray}}
\newcommand{\ee}{\end{equation}}
\newcommand{\ea}{\end{eqnarray}}

\title[The {\sl Swift}/UVOT view of NGC\,4321] {
The {\sl Swift}/UVOT catalogue of NGC\,4321 star forming sources:\\
A case against density wave theory}
\author[Ferreras et al.]  {Ignacio Ferreras$^1$\thanks{Email:
    ferreras@star.ucl.ac.uk}, Mark Cropper$^1$, Daisuke Kawata$^1$, 
  Mat Page$^1$, Erik~A. Hoversten$^2$\\
$^1$ Mullard Space Science Laboratory, University College London,
  Holmbury St Mary, Dorking, Surrey RH5 6NT\\
$^2$ Department of Astronomy and Astrophysics, The Pennsylvania State
  University, 525 Davey Laboratory, University Park, PA 16802, USA\\ }

\begin{document}
\date{\sl MNRAS in press.\\Accepted 2012 March 29.  Received 2012 March 9; in original
  form 2011 September 26}
\pagerange{\pageref{firstpage}--\pageref{lastpage}} \pubyear{2012}
\maketitle
\label{firstpage}

\begin{abstract}
  We study the star forming regions in the spiral galaxy NGC\,4321
  (M100). We take advantage of the spatial resolution (2.5\,arcsec
  FWHM) of the {\sl Swift}/UVOT camera and the availability of three
  UV passbands in the region $1600<\lambda <3000$\AA\ , in combination
  with optical and IR imaging from SDSS, KPNO/H$\alpha$ and {\sl
    Spitzer}/IRAC, to obtain a catalogue of 787 star forming regions
  out to three disc scale lengths. We use a large volume of star
  formation histories, combined with stellar population synthesis, to
  determine the properties of the young stellar component and its
  relationship with the spiral arms. The H$\alpha$ luminosities of the
  sources have a strong decreasing radial trend, suggesting more
  massive star forming regions in the central part of the galaxy. When
  segregated with respect to NUV-optical colour, blue sources have a
  significant excess of flux in the IR at 8$\mu$m, revealing the
  contribution from PAHs, although the overall reddening of these
  sources stays below E(B--V)=0.2~mag. The distribution of distances
  to the spiral arms is compared for subsamples selected according to
  H$\alpha$ luminosity, NUV--optical colour, or ages derived from a
  population synthesis model. An offset is expected between these
  subsamples as a function of radius if the pattern speed of the
  spiral arm were constant -- as predicted by classic density wave
  theory. No significant offsets are found, favouring instead a
  mechanism where the pattern speed has a radial dependence.
\end{abstract}


\section{Introduction}

Ninety years after the Great Debate on the the nature of ``spiral
nebulae'' \citep{grdeb}, our knowledge about disc galaxies has grown
considerably. However, the processes responsible for their spiral
appearance are still under scrutiny. Various mechanisms have been
proposed, the widely accepted one being the triggering of a long-lived
density wave that moves as a solid body, with constant pattern speed
\citep{lshu64}. One can define the corotation radius as the point at
which gas and stars move with the same speed as the spiral
pattern. Since the rotation velocity of gas and stars is roughly
independent of radius, the regions inside (outside) the corotation
radius will move faster (slower) than the spiral arm pattern. The gas
component piles up in the spiral arms, experiencing a shock that gives
rise to star formation \citep{rob69}.  One would expect in this
scenario that the youngest stars born from the molecular clouds in the
spiral arms would be found slightly ahead of the arm traced by the
molecular gas, if located within the corotation radius, and behind the
arm outside of corotation. Therefore, if we observe the tracers of the
different stages of star formation, such as HI, CO and H$\alpha$, we
expect to find an offset among them as a function of the radius. There
are a number of observational efforts to detect such offset in the
past \citep[e.g.][]{vog88,gb93,rand95}, and more recently
\citep[e.g.][]{tamb08,egu09,foy11}.  The majority of these studies
measure the angular offsets between CO as a tracer of star forming
molecular clouds and H$\alpha$ and/or $24\mu$m as a tracer of young
stars. A time difference around a few Myr is found \citep[e.g.][and
  references therein]{egu09} between CO and H$\alpha$. We note that HI
is not necessarily a direct tracer of molecular cloud formation, as it
often results from molecular gas dissociation caused by star formation
activity \citep{allen86,young91}.  A stronger constraint can be
imposed if we target instead tracers that cover a wider range of ages
for the distribution of recently formed sources. The use of UV light
opens up this possibility, since stellar populations can be tracked
over the first few hundred Myr \citep{foy11}.

An alternative mechanism for the triggering of spiral arms involves
self-gravitational instabilities \citep{glb65,toom81} whose spiral
features have much shorter lifetimes. Numerical simulations have not
been capable of explaining the onset of a stable density wave with
constant pattern speed, whereas short-lived spiral arms appear
naturally \citep[see
  e.g.][]{sellwood11,fuj11,quill11,wada11,rob12}. Other processes,
like tidal interactions or star-formation induced structures have been
proposed as well, leaving a different imprint on the properties of the
underlying stellar populations \citep{dp10}.

In this paper we make use of the Ultraviolet/Optical Telescope
\cite[UVOT,][]{uvot} on board the {\sl Swift} spacecraft
\citep{swift}, to study in detail the star forming regions of a nearby
grand-design spiral. NGC\,4321 (M100) is a face-on late-type spiral
galaxy (SABbc) towards the Virgo cluster \citep{deV91}.  Even though
{\sl Swift} is an observatory built for the study of transient
phenomena, most notably Gamma Ray Bursts, the resolution of UVOT makes
it an optimal instrument to study star formation in nearby
galaxies. The presence of a recent supernova in NGC\,4321
\citep[SN2006X,][]{sm06} triggered a number of follow-up observations
with {\sl Swift}/UVOT, allowing us to put together one of the deepest
UV images of a nearby galaxy at a spatial resolution of 2.5\,arcsec
(FWHM) over a $\sim 100$~arcmin$^2$ field of view.  In this paper we
adopt the distance from the Hubble Space Telescope Key Project to
measure Hubble's constant \citep{dm}, with a distance modulus
$\mu=30.91\pm 0.07$~mag, implying a distance of 15.2~Mpc, and an
angular scale of 73.7~pc/arcsec.  The scale length of the disc of
NGC\,4321 is h=1.39~arcmin or 6.15~kpc measured in the V band
\citep{kod86}. We use this value as a characteristic disc length
throughout this paper. The radial positions of the sources are
deprojected by using an inclination of 38$\degr$, with the receding
side located at a position angle of 151$\degr$ \citep{chemin06}.

\begin{table}
\caption{Log of {\sl Swift}/UVOT observations of NGC\,4321 used in this paper.}
\label{tab:obslog}
 \begin{tabular}{ccrrr}
\hline
 & & \multicolumn{3}{c}{Exposure Time (s)}\\
ObsID & Date & UVW2 & UVM2 & UVW1\\ 
 & & 1928\AA & 2246\AA & 2600\AA\\
\hline 
00035227001 & 2005/11/06 &  532.0 &    0   &   80.6\\
00035227002 & 2005/11/08 &  386.6 &  290.1 &  192.7\\
00035227003 & 2005/11/13 & 6832.3 & 5107.6 & 3414.5\\
00030365002 & 2006/02/08 &  673.6 &  505.6 &  332.2\\
00030365005 & 2006/02/10 & 1023.6 &    0   &  343.1\\
00030365006 & 2006/02/12 & 1092.1 &  390.6 &    0\\
00030365007 & 2006/02/14 &  939.6 &  704.6 &  462.1\\
00030365008 & 2006/02/17 &  998.0 &  745.0 &  486.2\\
00030365009 & 2006/02/18 &   17.8 &    0   &    0\\
00030365011 & 2006/02/20 &  595.4 &  298.5 &   98.4\\
00030365012 & 2006/02/21 &  325.6 &  244.6 &  160.9\\
00030365013 & 2006/02/24 &  543.2 &  405.2 &  266.8\\
00030365014 & 2006/02/26 &  519.2 &  387.2 &  253.0\\
00030365015 & 2006/02/28 &  769.2 &  578.2 &  377.9\\
00030365016 & 2006/03/02 &  443.4 &  332.4 &  159.9\\
00030365017 & 2006/03/05 &  701.3 &  505.6 &  331.2\\
00030365018 & 2006/03/12 & 1032.4 &  773.4 &  508.2\\
00030365019 & 2006/03/18 & 1103.4 &  827.4 &  542.7\\
00030365021 & 2006/03/27 & 1045.7 &    0   &  468.1\\
00030365023 & 2006/03/29 & 1780.2 &    0   & 1368.0\\
\hline
 & TOTAL (ks) & 21.35 & 12.10 & 9.85\\
 & Effective TOTAL (ks) & 21.28 & 11.95 & 9.74 \\
\hline
\end{tabular}
\end{table}

\section{The UV images}

We extracted the UV images of NGC\,4321 from the archive of the {\sl
  Swift}/UVOT mission\footnote{\tt http://heasarc.nasa.gov/W3Browse/}
at HEASARC. The log of the observations can be found in
Table~\ref{tab:obslog}. UVOT is one of the three instruments on board
the {\sl Swift} observatory. It couples a 30~cm Ritchey-Chr\'etien
telescope to a photon-counting, microchannel-plate-intensified CCD
camera. Details of the instrument can be found in \citet{uvot},
whereas the calibration is described in \citet{poole08} and
\cite{bree10,bree11}.  We use the three available UV filters: UVW2,
UVM2 and UVW1, that span the wavelength range: 1600-3400\AA\ . These
three filters roughly map the spectral window of the {\sl GALEX} NUV
passband \citep{galex}, with UVM2 centered at the position of the
2175\AA\ UV-bump in the Milky Way extinction curve
\citep{con10}. Fig.~10 in \citet{poole08} shows the in-orbit effective
area of the UVOT filters. The spatial resolution of the instrument,
along with a Point Spread Function of 2.4, 2.5 and 2.9\,arcsec (FWHM)
in UVW2, UVM2 and UVW1, respectively \citep{bree10}, makes UVOT images
significantly better resolved than NUV images with {\sl GALEX}, which
has a FWHM of $\sim$6\,arcsec \citep{morrissey05}. We note that at the
distance of M100, the UVOT resolution limit maps a physical distance
around 200~pc, which is the maximum possible value to be able to
neglect deblending issues in HII regions, given their characteristic
sizes \citep{ken89}.

\begin{figure*}
  \includegraphics[width=15cm]{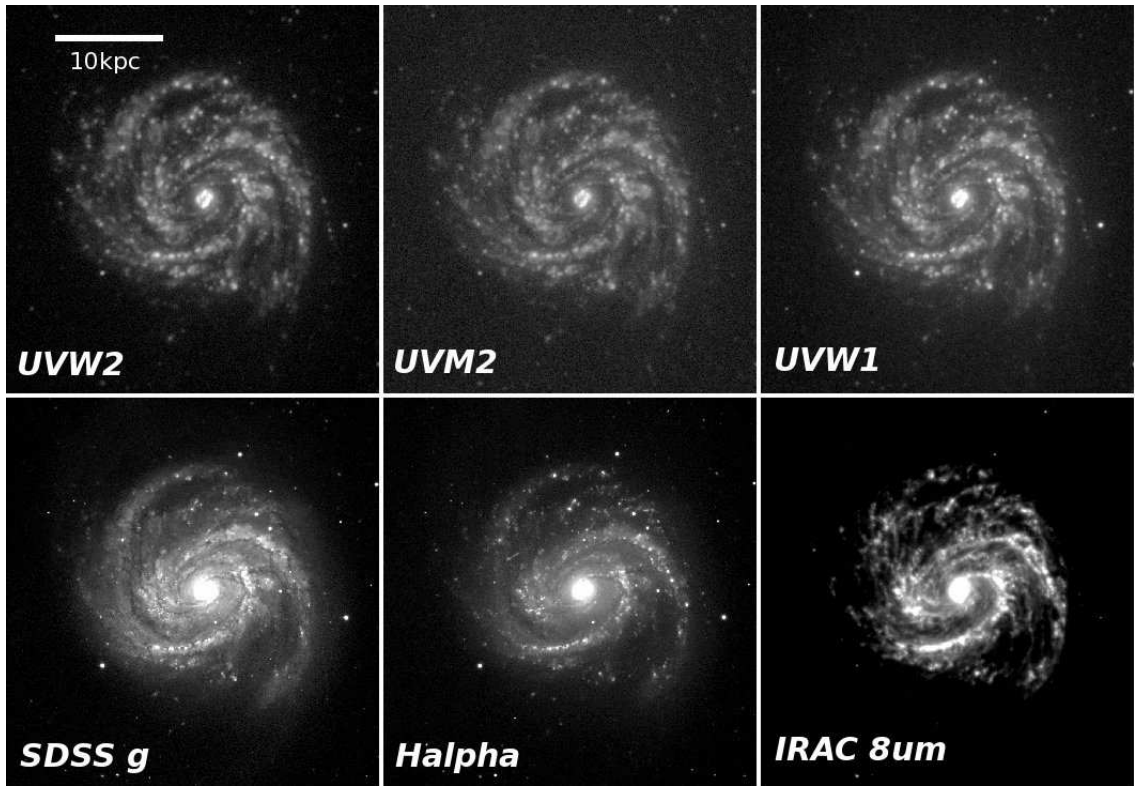}
  \caption{Views of NGC\,4321 (M100) in different passbands over a
    $8\times 8$ arcmin$^2$ field of view. North is up, East to the
    left. The top panels show deep NUV imaging with the {\sl
      Swift}/UVOT instrument. The bottom panels show optical/IR images
    from archival data, as labelled.}
  \label{fig:views}
\end{figure*}

\begin{figure*}
 \includegraphics[width=16.5cm]{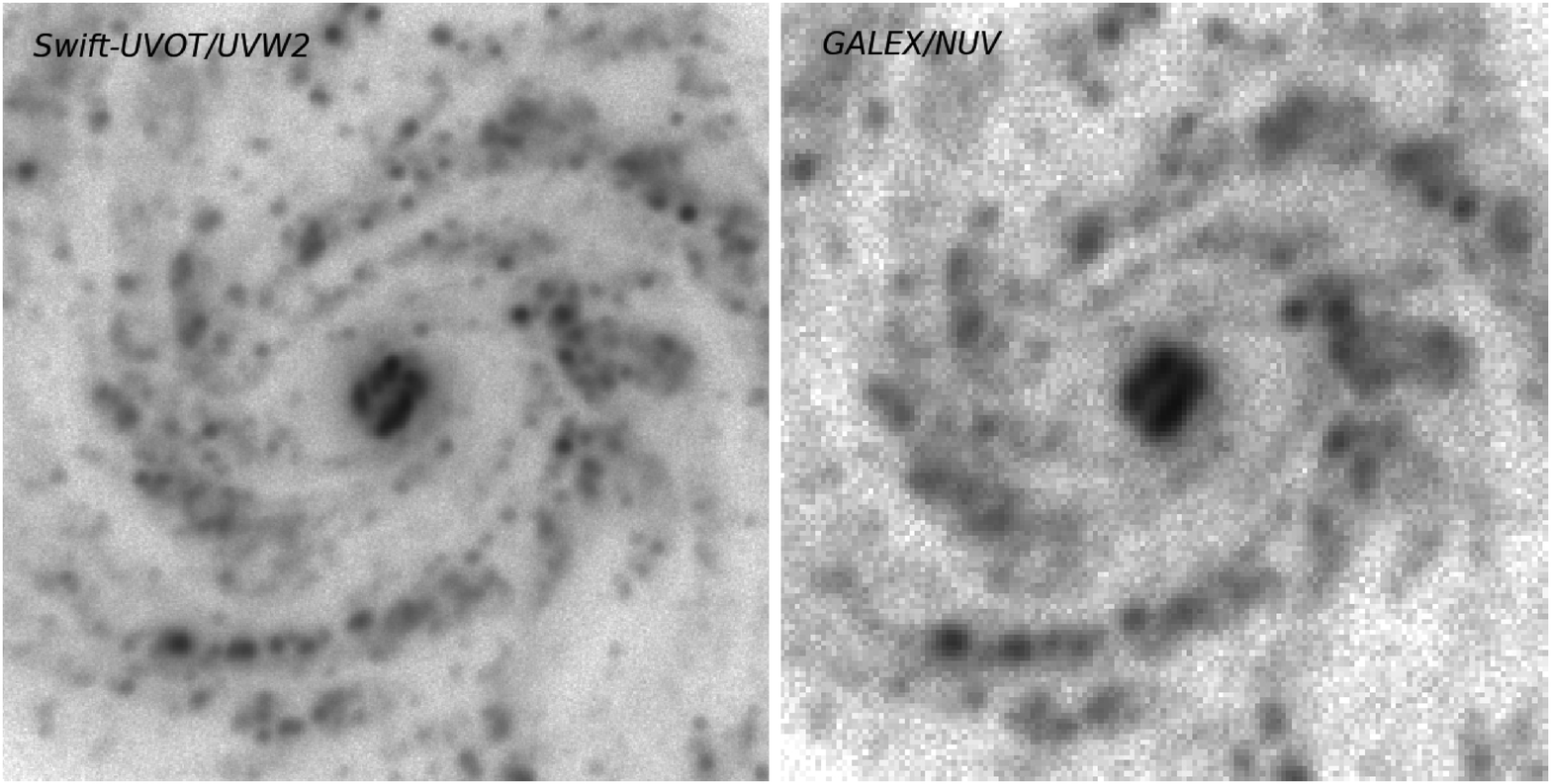}
 \caption{A comparison of the depth and resolution between our
   UV imaging (UVOT/UVW2, 21.3ks, left), and the available archival
   image from {\sl GALEX}/NUV (1.1ks, right). Both frames show the central
   $8\times 8$~kpc$^2$ region.}
 \label{fig:galex}
\end{figure*}

We retrieved the science-grade exposures from the archive.  Some of
these images did not have an aspect correction. This failure is often
caused by the lack of nearby stars in the USNO catalogue, that
are used as reference in an automatic way to determine this
correction. In order to have as many images as possible, we wrote our
own software to aspect correct the images from a manual selection of
foreground stars, and to combine all images with a flux-conserving
drizzling algorithm at the same pixel size (i.e. 0.5\,arcsec), based
on the method of \citet{fh02}. Some of the images, with exposure
times typically below 30 seconds, could not be used because they could
not be properly registered. Hence, we define as effective integration
times those from the final exposure map within the footprint of
NGC\,4321. The exposure map in this region is homogeneous.  The total
effective exposure times of the UV data are 21.28ks in UVW2, 11.95ks
in UVM2 and 9.74ks in UVW1. The top panels of Fig.~\ref{fig:views}
shows greyscale images of the UVOT combined exposures. We illustrate
in Fig.~\ref{fig:galex} the difference between our UVOT/UVW2 deep
exposure and the available archival image from {\sl GALEX}/NUV
(exposure time 1183 s). Both frames show the central
$8\times8$~kpc$^2$ region.

\begin{figure*}
 \includegraphics[width=16.5cm]{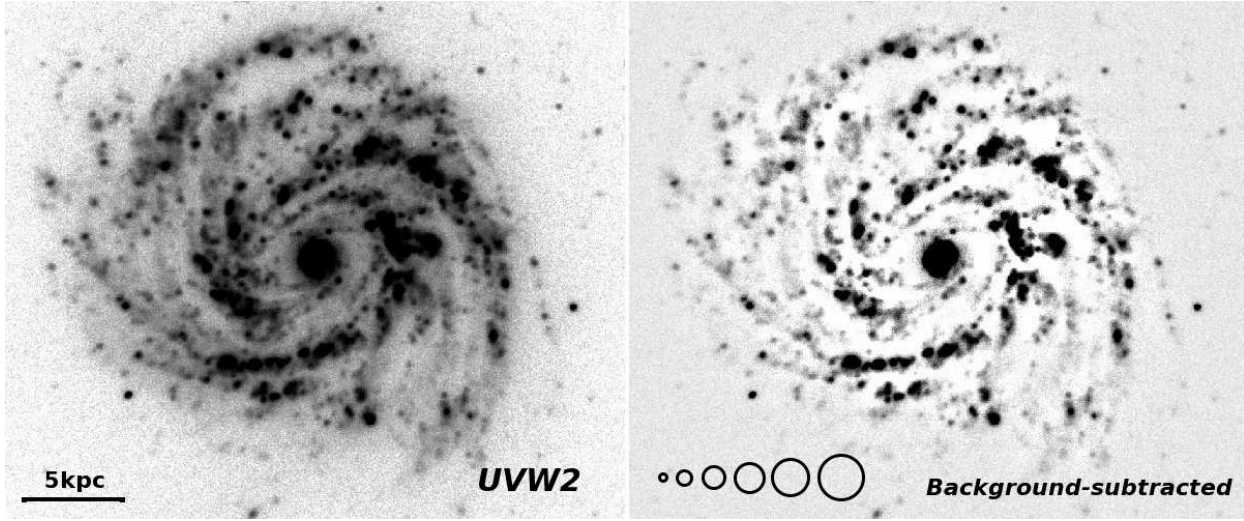}
 \caption{Illustration of our background subtraction method. The scale
   is the same for both images. The circles are the apertures used for
   the selection of the background at a given point (see text for
   details).}
 \label{fig:back}
\end{figure*}

\begin{figure}
  \includegraphics[width=8.5cm]{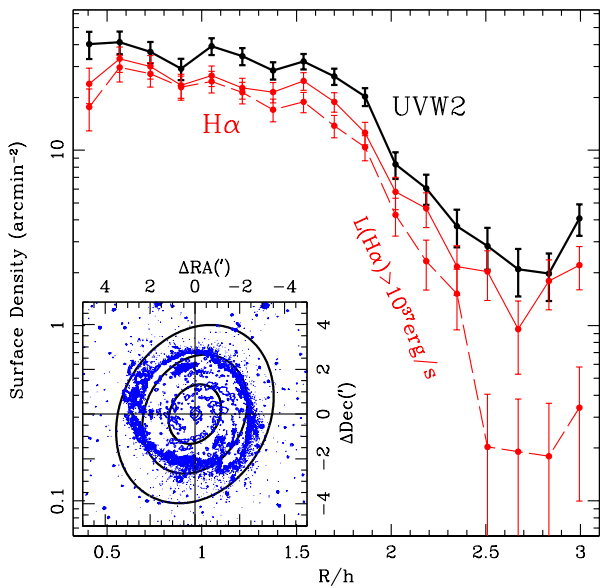}
  \caption{Radial profile of the surface density of UVW2 sources
    (thick black line), including the density of counterparts in the
    H$\alpha$ images (red).  The surface density of bright H$\alpha$
    sources (L(H$\alpha$)$>10^{37}$erg/s) is shown as a dashed red
    line. The error bar give the Poisson fluctuations. The inset shows
    the extent of the UVW2 disc, with projected circles -- according to
    the orientation of the disc -- representing the R=h, 2h, 3h radii,
    where h is the characteristic scale-length measured in the V-band
    \citep{kod86}}
  \label{fig:Ncounts}
\end{figure}

The source detection was performed with SExtractor \citep{sex} on the
UVW2 image, which is the deepest one of the three, and corresponds to
the shortest wavelength -- optimally mapping ongoing star forming
regions. The SExtractor code has been optimized for the detection of
sources over a homogeneous background (e.g. galaxies in a survey or
stars in a globular cluster). However, our case is slightly different,
as the star forming sources are located on top of a diffuse background
that has to be removed before detecting the sources. To contend
with this, we define our own background on the original UVW2 image,
following a procedure inspired by the methodology set out in
\citet{hovers11}.  A number of images are obtained by performing a
circle median of the observed image over a range of radii. In order to
determine the background, we need to subtract the information over
smaller scales, ranging from individual HII regions to structures
generated by the superposition of clustered, unresolved star forming
sources \citep[see][for a detailed analysis of the effects of poorer
  spatial resolution]{pleuss00}. The spatial resolution (2.5\,arcsec),
pixel size (0.50 arcsec) and galaxy scale (74 pc/arcsec) of the images
led us to choose the following radii for the median kernel:
R=\{5,10,15,20,25,30\} pixels, which map into a physical scale between
0.18 and 1.11\,kpc.  The smallest option maps scales over the largest
sizes of HII regions \citep{sco01}. We combine all those images along
with the original one to determine, pixel by pixel, the background as
the {\sl minimum} pixel value among these images. Fig.~\ref{fig:back}
compares the UVW2 image with the background-subtracted one following
this methodology.  The selection criteria applied to SEXtractor,
impose a detection threshold at a minimum of 9 connected pixels with a
count rate $2\sigma$ above the background. The final catalogue
comprises 787 sources out to R=3h. Fig.~\ref{fig:Ncounts} shows the
radial trend of the number density of NUV-detected sources, and those
with available H$\alpha$ photometry -- which amount to 68\% of the
total sample.  Notice the level in the number counts around R=2h (see
also inset of Fig.~\ref{fig:Ncounts}), revealing a sharp truncation of
the UV light tracing the youngest stellar populations in NGC\,4321.
The brightest H$\alpha$ sources, corresponding to stronger sites of
star formation, are preferentially located in the central regions. We
will discuss this issue in more detail in \S4. The positions of the
detected sources are used for the photometric analysis. Note the
photometry is performed on the original image, not the
background-subtracted one. Hence, the photometry can be described as a
superposition of a young, star forming region, and an older component
from the background populations.

Since UVOT is a photon counter, its photometric accuracy for brighter
sources is limited by the sampling rate of the CCD, which is 91 frames
per second. Bright sources will experience a coincidence loss effect,
whereby the true flux is underestimated as all photons hitting the
detector in a single frame and within a small region, are counted as
one. This effect has been properly characterized in \citet{poole08}
and \citet{bree10}, and is especially important with the optical
passbands, where the photon count rate is much higher \citep[as
  opposed to the UV, see Fig.~1 of][to illustrate this point]{hovers11}.

\begin{table*}
\caption{Photometry of UVW2 selected sources in
NGC\,4321$^1$: {\sl Swift}/UVOT and SDSS (optical).}
\vbox to220mm{
  \includegraphics[width=15cm]{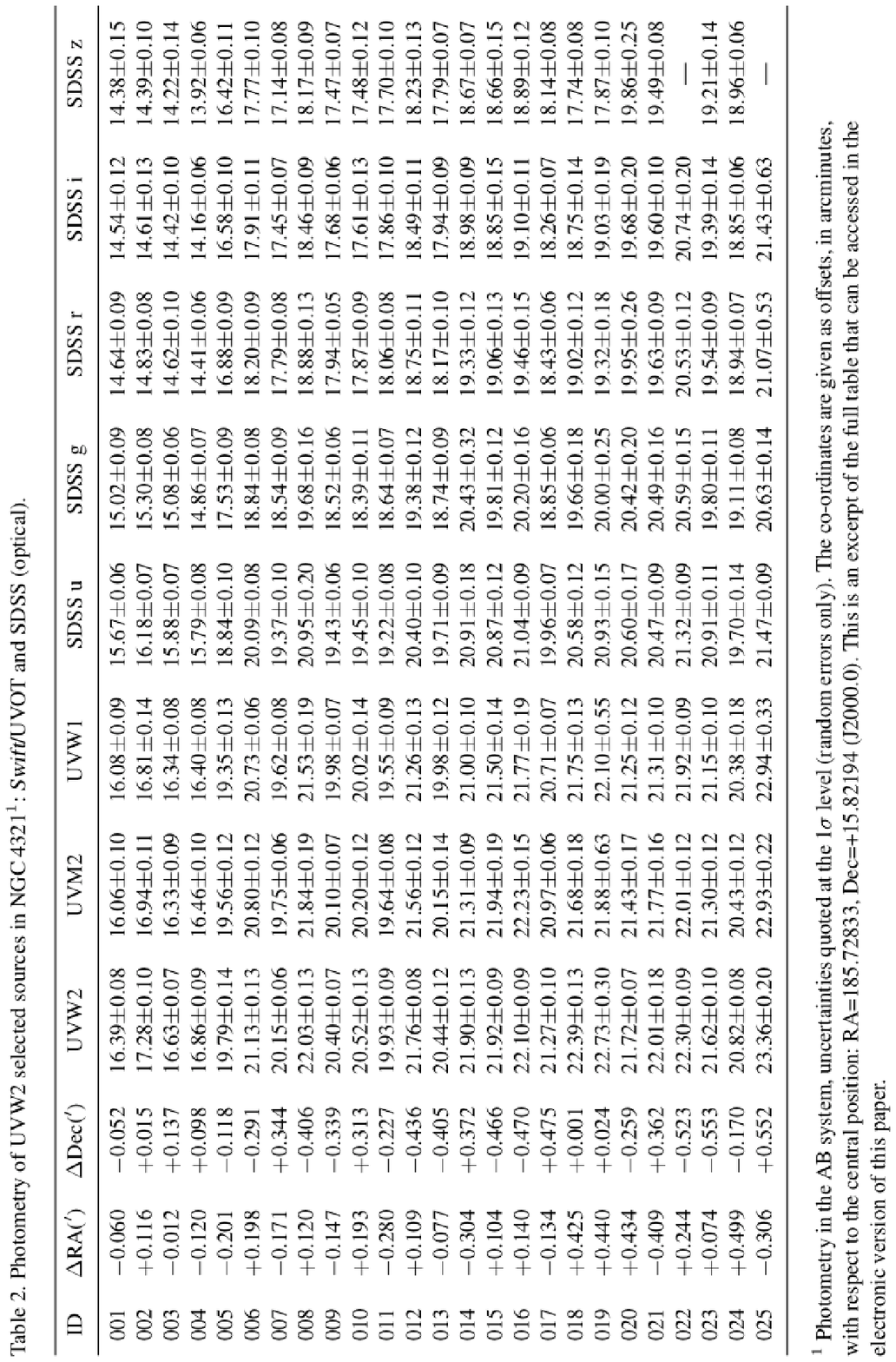}
  \label{tab:photo1}
}
\end{table*}

\begin{figure}
  \includegraphics[width=8.5cm]{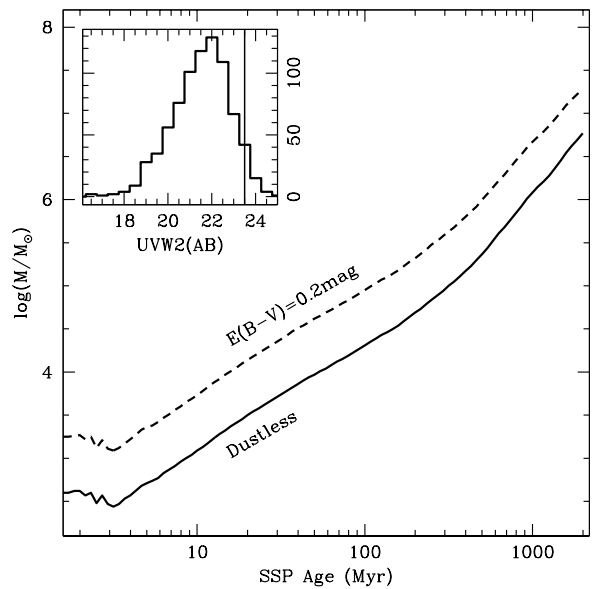}
  \caption{The stellar mass is shown with respect to the age of a
    simple stellar population for the limiting magnitude of UVW2=23.5 (at
    S/N$\sim$5). Dustless (solid)
    and a dusty case (dashed) is considered, as labelled. The inset
    shows the histogram of UVW2 detections, with the vertical line
    locating the S/N$\sim 5$ case.}
  \label{fig:w2limit}
\end{figure}

In our case, optical photometry is available from the Sloan Digital
Sky Survey (see \S\ref{sec:pan}), so only the UV data from UVOT is
used. For the UV images we apply the corrections defined in
\citet{poole08}. Our photometric measurements are performed within a
R=3\,arcsec aperture (see \S3). For the integration times of the final
images, the {\sl Swift}/UVOT exposure time calculator\footnote{\tt
  http://www.mssl.ucl.ac.uk/www\_astro/uvot} gives point-source
detections at S/N$\sim 5$ of UVW2=23.6, UVM2=22.80 and UVW1=22.7 (AB),
in good agreement with our photometric data: Using the detection of a
$5\sigma$ fluctuation above the background within the extent of a PSF
FWHM, we obtain limiting magnitudes in the final images of UVW2=23.5,
UVM2=23.2, and UVW1=22.9 (AB).  Fig.~\ref{fig:w2limit} translates the
UVW2=23.5 limit into a stellar mass limit as a function of age --
according to the simple stellar populations from the 2009 version of
the \citet{bc03} models for a \cite{chab03} Initial Mass
Function. Both dustless and a dusty, E(B--V)=0.2 mag model are
considered. The inset shows the distribution of UVW2 sources, with
a vertical line marking the S/N=5 detection limit. A comparison of
photometric measurements between {\sl GALEX}/NUV (see
Fig.~\ref{fig:galex}) and UVOT/UVM2 -- which is the filter that tracks
best the throughput of the NUV passband -- gives good agreement, with
an RMS scatter for the difference NUV$-$UVM2 around 0.18\,mag (taking
sources brighter than UVM2=22AB). This scatter is mainly dominated by
the fact that the measurements are taken over a crowded field.

\begin{table*}
\begin{minipage}{115mm}
\caption{Photometry of UVW2 selected sources in NGC\,4321$^1$:
  Spitzer/IRAC and H$\alpha$.}
\label{tab:photo2}
\begin{tabular}{ccccccr}
\hline
ID & $R/h$ & [$3.6\mu$m] & [$4.5\mu$m] & [$5.8\mu$m] & [$8\mu$m] & log H$\alpha^\dag$\\
\hline 
001 & 0.072 &  13.19$\pm$0.09 & 13.06$\pm$0.10 & 12.27$\pm$0.07 & 11.05$\pm$0.07 &  2.31$\pm$0.04\\
002 & 0.103 &  13.02$\pm$0.12 & 12.91$\pm$0.11 & 12.05$\pm$0.14 & 10.81$\pm$0.15 &  2.06$\pm$0.07\\
003 & 0.104 &  13.02$\pm$0.10 & 12.89$\pm$0.09 & 12.23$\pm$0.07 & 11.02$\pm$0.05 &  2.44$\pm$0.03\\
004 & 0.116 &  12.68$\pm$0.07 & 12.59$\pm$0.08 & 11.88$\pm$0.06 & 10.68$\pm$0.06 &  2.16$\pm$0.05\\
005 & 0.212 &  15.50$\pm$0.12 & 15.47$\pm$0.10 & 15.08$\pm$0.09 & 13.99$\pm$0.07 &    ---   \\
006 & 0.254 &  16.96$\pm$0.13 & 16.97$\pm$0.10 & 16.95$\pm$0.12 & 16.60$\pm$0.22 &  0.26$\pm$0.08\\
007 & 0.276 &  15.79$\pm$0.08 & 15.60$\pm$0.06 & 14.76$\pm$0.09 & 13.49$\pm$0.06 &  1.32$\pm$0.07\\
008 & 0.309 &  17.25$\pm$0.07 & 17.30$\pm$0.08 & 16.90$\pm$0.07 & 15.81$\pm$0.09 &  0.24$\pm$0.33\\
009 & 0.312 &  16.34$\pm$0.09 & 16.29$\pm$0.07 & 15.40$\pm$0.07 & 14.23$\pm$0.09 &  0.94$\pm$0.07\\
010 & 0.320 &  16.58$\pm$0.14 & 16.44$\pm$0.13 & 15.98$\pm$0.10 & 14.86$\pm$0.11 &  0.86$\pm$0.08\\
011 & 0.327 &  16.46$\pm$0.10 & 16.33$\pm$0.08 & 15.12$\pm$0.08 & 14.00$\pm$0.08 &  1.25$\pm$0.06\\
012 & 0.330 &  17.38$\pm$0.09 & 17.44$\pm$0.08 & 17.11$\pm$0.10 & 15.90$\pm$0.09 &  0.34$\pm$0.07\\
013 & 0.331 &  16.80$\pm$0.08 & 16.75$\pm$0.09 & 16.22$\pm$0.08 & 15.08$\pm$0.07 &  0.87$\pm$0.03\\
014 & 0.349 &  17.03$\pm$0.13 & 16.79$\pm$0.11 & 15.61$\pm$0.12 & 14.31$\pm$0.09 &  1.16$\pm$0.03\\
015 & 0.352 &  17.80$\pm$0.14 & 17.85$\pm$0.09 & 17.42$\pm$0.11 & 16.24$\pm$0.11 & -0.04$\pm$0.28\\
016 & 0.358 &  17.82$\pm$0.13 & 17.80$\pm$0.14 & 17.19$\pm$0.15 & 15.76$\pm$0.17 & -0.00$\pm$0.22\\
017 & 0.361 &  17.42$\pm$0.14 & 17.42$\pm$0.16 & 18.18$\pm$0.88 & 20.61$\pm$1.19 & -0.96$\pm$0.08\\
018 & 0.370 &  17.97$\pm$0.11 & 17.93$\pm$0.15 & 18.67$\pm$0.62 &    ---    &  0.70$\pm$0.06\\
019 & 0.387 &  18.35$\pm$0.23 & 18.40$\pm$0.17 & 19.34$\pm$0.68 &    ---    &  0.74$\pm$0.04\\
020 & 0.391 &  17.81$\pm$0.15 & 17.46$\pm$0.16 & 16.77$\pm$0.16 & 15.77$\pm$0.59 &  0.85$\pm$0.17\\
021 & 0.406 &  17.33$\pm$0.10 & 17.03$\pm$0.06 & 15.40$\pm$0.09 & 14.10$\pm$0.08 &  1.25$\pm$0.08\\
022 & 0.415 &     ---    &    ---    & 18.78$\pm$0.43 & 16.55$\pm$0.45 &    ---   \\
023 & 0.417 &  18.56$\pm$0.22 & 18.81$\pm$0.28 &    ---    &    ---    &    ---   \\
024 & 0.428 &  17.75$\pm$0.12 & 17.56$\pm$0.12 & 16.45$\pm$0.14 & 15.22$\pm$0.17 &  0.82$\pm$0.06\\
\hline
\end{tabular}
\medskip $^1$Photometry for the IRAC bands in the AB system,
uncertainties quoted at the 1$\sigma$ level (random errors only). This
is an excerpt of the full table that can be accessed in the electronic
version of this paper.\\ $^\dag$ measured as
$\log_{10}[L(H\alpha)/10^{37}{\rm erg/s}]$.
\end{minipage}
\end{table*}

\section{A panchromatic view of NGC\,4321}
\label{sec:pan}

In addition to NUV data, we retrieved the optical images from the
Sloan Digital Sky Survey/DR8 \citep{sdssdr8}. NGC\,4321 falls at the
intersection of four different SDSS fields (run=3631, field=456,457;
and run=4381, field=99,100). We used the mosaic making facility from
the SDSS-III Service Archive Server (http://data.sdss3.org/mosaics) to
retrieve a properly calibrated image in all five bands ($u,g,r,i,z$)
and registered the images. In addition, imaging is available from
Spitzer/IRAC, and KPNO (H$\alpha$). The reduced, calibrated images are
taken from the Fifth data delivery (April 2007) of the SINGS
catalogue\footnote{\tt http://sings.stsci.edu} \citep{sings}. All the
optical and IR images are re-sampled to a common size and pixel scale,
given by the UVW2 frame, which is the image used for the selection of
targets. We use the IRAF task {\sl wregister} to ensure all bands are
aligned properly and with the UVOT original pixel size
(0.502\,arcsec). We compared the photometric estimates before and
after this transformation, to find no significant systematic
change. Fig.~\ref{fig:views} shows some of these frames, from the
  UV to the IR.

The photometry for all images (UV, optical and IR) is performed with
our own code, described below, motivated by the need for a consistent
background subtraction and error estimate. For each position,
determined by the SExtractor run, we measure the flux within a
R=3\,arcsec aperture. At every pixel the background is estimated
within a circular region with radius R=20\,arcsec, where all pixels
already allocated to a source by SExtractor (i.e. on the segmentation
map) are removed. This distribution is then clipped at the 3$\sigma$
level and the median of the resulting set is taken as the
background. The IRAC measurements include standard aperture
corrections as prescribed in the instrument
handbook\footnote{http://irsa.ipac.caltech.edu/data/SPITZER/docs/irac/iracinstrumenthandbook/}.
We note that the standard calibration of the UVOT instrument is done
on photometric measurements of point sources within a R=5\,arcsec
aperture \citep{poole08,bree10}. Given the spatial resolution of the
camera, this choice of aperture is optimal for the photometry of Gamma
Ray Bursts -- the main target of UVOT science. However, it does not
take advantage of the spatial resolution in this case. Hence, we chose
a smaller radius for the measurements (R=3\,arcsec), and applied an
aperture correction to align with the standard calibration. This
correction is obtained by performing aperture photometry on a number
of isolated stars in the frame. We selected five bright unresolved
targets, for which the photometry is compared between a R=3\,arcsec
and a R=5\,arcsec aperture. This result is extended to the SDSS, and
NOAO/H$\alpha$ frames, to correct for the difference in PSF between
the different images. For the {\sl Spitzer}/IRAC photometry, we
already make use of the published aperture corrections, as explained
above.  The corrections for the UVW2, UVM2, UVW1 filters are $0.150$,
$0.215$, and $0.168$~mag, respectively. In contrast, the aperture
corrections for the SDSS images are $0.030$, $0.045$, $0.092$,
$0.073$, and $0.052$~mag, in the u, g, r, i, and z bands,
respectively. Finally, the NOAO H$\alpha$ image needs a correction
between these apertures of $\Delta\log$H$\alpha=0.126$~dex.  Tables~2
and 3 show an excerpt of the aperture corrected photometry for
the sources in all {\sl Swift}/UVOT, SDSS, KPNO/H$\alpha$ and
Spitzer/IRAC bands, with the error bars quoted at the 1$\sigma$
level. The photometry in the tables is corrected for (Milky Way)
foreground extinction according to the reddening law of \citet{fitz99}
for E(B--V)=0.026 towards the direction of NGC\,4321, taken from the
maps of \citet{schleg98}.

\section{Modelling the stellar populations}

\begin{figure}
  \includegraphics[width=8.5cm]{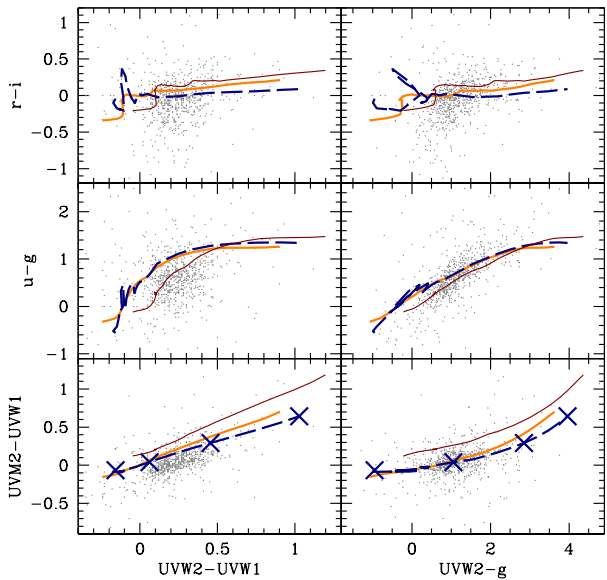}
  \caption{Colour-colour diagrams of the star forming regions in
    NGC\,4321. The observed magnitudes have already been corrected for
    foreground extinction (E(B--V)=0.026). The models correspond to
    dustless CB09 (thick solid line, orange), SB99 (dashed line, blue)
    and CB09 with a E(B--V)=0.2~mag dust screen (thin solid, red). See
    text for details. The SB99 models in the bottom panels are marked
    by crosses for ages (from left to right) of 1, 100, 500 and
    900~Myr. }
  \label{fig:CCD}
\end{figure}

Fig.~\ref{fig:CCD} shows the NUV and optical colours of the
UVW2-selected sources. A number of colour-colour diagrams are shown,
involving UVOT and SDSS photometry. The observations (dots) match the
predictions for a set of simple stellar populations (SSP). All models
have solar metallicity. The thick solid orange lines correspond to
dustless SSPs from the latest models of \citet{bc03}, with a
\citet{chab03} initial mass function (IMF). To illustrate the effect
of using a different population synthesis model, or a different IMF,
we also show in the figure the colour tracks from the Starburst99
dustless models \citep{sb99} with a \citet{salp55} IMF (blue dashed
lines). For reference, crosses mark stellar ages of 1, 100, 500 and
900~Myr in the latter. We also include, as a thin solid red line, the
effect of a E(B--V)=0.2~mag dust screen on the Bruzual \& Charlot
models, according to a Milky Way reddening law \citep{fitz99}. The use
of this reddening law is justified by the type of dust present in
NGC\,4321 \citep[see e.g.][]{draine07}. The model predictions clearly
do not favour large amounts of dust, as expected from a UV-based
source detection. Notice that on the colour-colour diagrams the dust
content of the sources is found within a small range of E(B--V)$\simlt
0.2$\,mag. We will quantify this statement in some detail
below. Fig.~\ref{fig:CCD} also reveals that the age of these sources
is never older than about 1~Gyr, although we have to take care of the
presence of background populations of older stars.  Note the UV
colours are in better agreement with the SSP models, whereas the
optical colours are more dispersed. This is caused by the contribution
at redder wavelengths from older populations within the apertures. We
explore this issue in more detail in the next figure.

\begin{figure}
  \includegraphics[width=8.5cm]{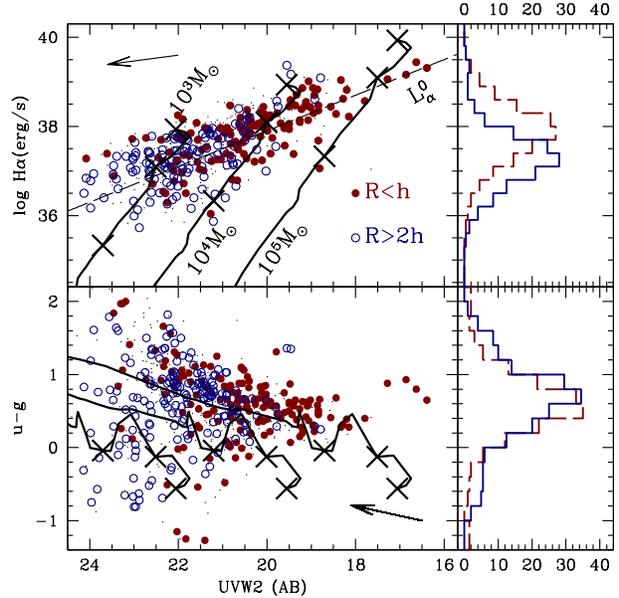}
  \caption{The H$\alpha$ line luminosities and optical colours of the
    star forming regions are shown with respect to UVW2
    magnitudes. The star forming regions are split with respect to
    galactocentric distance, as labelled. The lines track the
    predictions from Starburst99 \citep{sb99} of a simple stellar
    population with solar metallicity, and stellar masses between
    $10^3$ and $10^5$M$_\odot$. The crosses mark (from top to bottom)
    ages of 1, 5 and 10 Myr. The panels on the right show the
    distribution of inner (dashed) and outer (solid) star forming
    regions. The arrows in both panels correspond to a dust
    attenuation of E(B--V)=0.2~mag using a Milky Way reddening
    law. The dashed line in the top panel marks a limiting line
    luminosity $\log$\,L$_\alpha^0$(erg/s)$\equiv 37.9-0.4($UVW2$_{\rm AB}-20)$, used as
    a proxy to segregate H$\alpha$ detected sources with respect to
    age (see Fig.~\ref{fig:DensWave}).}
  \label{fig:Rad}
\end{figure}

Fig.~\ref{fig:Rad} shows the trend of the H$\alpha$ luminosities with
respect to UVW2 magnitude, for those sources with a detection in the
H$\alpha$ image (68\% of the total sample). H$\alpha$ emission in star
forming regions originates from gas excited by ionising photons from
OB stars, i.e. more massive than $\sim 15$M$_\odot$. In contrast, NUV
emission will also originate from the photospheres of lower mass
stars, down to $\sim$3M$_\odot$. Hence, the ratio between H$\alpha$
and NUV light works as a stellar clock over the first few million
years of the nascent population \citep[assuming a universal initial
  mass function, as expected for a bright galaxy, see][]{hg08}. The
solid lines in the top panel of Fig.~\ref{fig:Rad} correspond to the
model predictions from Starburst99 for a simple stellar population at
solar metallicity, with a Kroupa IMF \citep[see][and references
  therein]{sb99}. The Kroupa IMF is very similar to the Chabrier IMF
used below for the analysis of the photometry: differences between
them at the level of the data used are indistinguishable and do not
affect our conclusions. The three solid lines represent three
different stellar masses, from 10$^3$ to 10$^5$M$_\odot$. As age
progresses, the H$\alpha$ flux decreases: the crosses, from top to
bottom, mark the ages of 1, 5 and 10~Myr. The star forming sources are
coded according to radial position. To highlight the differences with
respect to galactocentric distance, we show those sources with R$<$h
(solid red dots) or R$>$2h (open blue dots). The radial trend is
evident, with the central sources being more massive, a result
consistent with the findings of \citet{godd11}. Notice the orientation
of the ``dust vector'' for E(B--V)=0.2\,mag. If dust were an important
factor in the inner regions (i.e. the solid red dots), the
segregation between central and outer sources in this graph would be
more prominent, as a reddening correction would shift the points
corresponding to the central regions towards the right of the figure,
therefore increasing the separation between inner and outer sources.
The ages of all sources are mostly younger than 10~Myr, as expected
for an H$\alpha$ detection. The bottom panels show the optical
colours, with the solid lines representing the Starburst99 predictions
for the same models as in the top panels. Differences with respect to
the radial distance are less evident, with a tail towards bluer
colours in the outer sources. Those sources with colours redder than
those from the simple stellar populations reflect the contribution
from an underlying older component, which will dominate optical and
NIR light. This figure illustrates the shortcomings of simple stellar
populations in describing the photometric data.

The simplest model to overcome this problem involves a superposition
of two SSPs. These models have been applied for the detection of
residual star formation over the old populations found in early-type
galaxies \citep[see e.g.][]{fs00,kav07}. The models we use in this
paper are described by five free parameters comprising the age of the
old and young components (t$_{\rm O}$ and t$_{\rm Y}$, respectively),
the metallicity of both populations (assumed to be the same), the mass
fraction in young stars (f$_{\rm Y}$) and the dust reddening affecting
the young component (E$_{\rm Y}$), whereas the old component is
considered to be dustless. In order to determine robust best fits, we
opted for a full search of this five dimensional parameter space. The
grid of parameters is shown in Table~\ref{tab:2bst}, involving over
100 million models. We compute the flux from the SSPs, integrated over
the filter response functions, and store them in a lookup table, to
maximise the computing speed of the fitting algorithm, which is also
parallelized to reduce wall-clock computing time. The effect of dust
on the magnitudes is modelled by a fourth order polynomial in E(B--V),
where the coefficients are computed from a grid of 16,384 SSP models
over a range of ages, metallicities and E(B--V), using a \citet{fitz99}
reddening law. For each choice of parameters, the photometric data
from the UVOT, SDSS and IRAC channels 1 (central wavelength
$\lambda_c=3.6\mu$m) and 2 ($\lambda_c=4.5\mu$m) are used to define a
likelihood. Following a Bayesian approach, we apply a Gaussian
prior on the metallicity, with mean $\log Z/Z_\odot=-0.1$ and standard
deviation 0.3~dex. The choice of mean is based on the observations of
the gas-phase metallicity of M100 \citep{mous10}. However, we note
that by removing the prior altogether, the age distribution of the old
and young components is not affected within the scatter found among
the sources.

\begin{figure}
  \includegraphics[width=8.8cm]{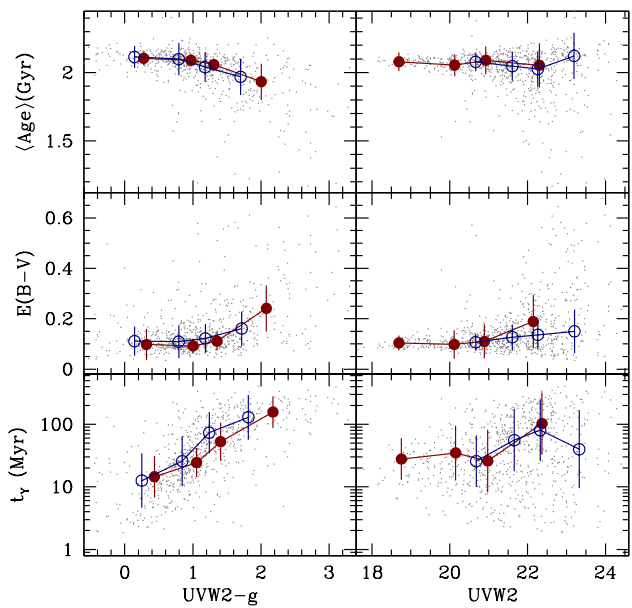}
  \caption{Distribution of ages and dust reddening of the UV sources
    according to the two-burst model (see text for details). The solid
    (open) dots show median values for subsamples at R$<$h (R$>$2h),
    with the binning chosen at equal number of sources per bin. The
    error bars correspond to the RMS scatter of individual sources
    within each bin. The top panels show the mass-weighted average
    age, whereas the bottom panels correspond to the age of the
    youngest component in the two-burst model adopted here (see text
    for details).}
  \label{fig:Ages}
\end{figure}

Fig.~\ref{fig:Ages} shows the distribution of the ages -- total mass
weighted average ({\sl top}) or age of the youngest component ({\sl
  bottom}) -- and dust reddening for all sources with respect to
UV-optical colour or UV flux. The big dots track the median of the
distribution within bins keeping equal number of sources per bin, with
the error bars representing the RMS scatter. Solid (open) dots
correspond to sources at radial distance from the centre R$<$h
(R$>$2h). As expected, UV-optical colour tracks the age of the young
component and no significant trend is found with respect to the dust,
obtaining a value for the reddening around E(B--V)$\simlt
0.2$~mag. The distribution of metallicities has a mean of $\log
Z/Z_\odot=-0.18$ -- as expected, slightly lower than the measured
gas-phase metallicity -- and an RMS scatter of $0.08$\,dex. There is
no significant radial trend in metallicity, although the RMS scatter
is higher at R$>$2h ($0.12$\,dex). Note that the ages of the young
component found here are significantly older than those estimated from
the H$\alpha$ luminosities (Fig.~\ref{fig:Rad}), reflecting the
complex mixture of populations within a resolution element \citep[see
  e.g.][]{pleuss00}.  Within a UVOT detection, we expect to have a
wide range of stellar populations, including dust obscured regions,
H$\alpha$ emitting sources, and more mature UV-bright areas. An
analysis based on the broadband photometry will be inherently biased
towards the last component. The presence of dust obscured regions can
be illustrated by the stacking of sources into two representative
UV-to-IR spectral energy distributions (SEDs). Fig.~\ref{fig:SEDs}
shows the photometric SEDs of the youngest sources (using the UVW2--g
colour as a proxy for age) have an IR excess in the IRAC bands, which
reflects the contribution from dust-obscured regions that are possibly
too faint in the UV-optical range to have an effect on the analysis of
the UVOT/SDSS photometry. However, this figure gives further support
to the use of UVW2--g as a proxy for the stellar ages of these
sources. Notice in Fig.~\ref{fig:Ages} the lack of any trend in the
properties of the stellar populations with respect to radial distance,
except for the fact that the outer sources are fainter in UV (see also
Fig.~\ref{fig:Rad}), confirming that the radial change refers to the
stellar mass content of the sources. The young component has a
distribution of ages in the range 10--100~Myr. We will use this
information in the next section to explore in detail the distribution
of UV sources with respect to the spiral arms.

\section{Putting density wave theory to the test}

The properties of the underlying stellar populations of the star
forming sources provide valuable constraints on the origin of spiral
arms. In the previous section, we showed that our NUV-detected sources
span a wide range of stellar ages (a few hundreds of Myr) and are
distributed over a large range of galactocentric radii. Therefore,
they represent an ideal tracer for the ``offset method'' \citep[see
  e.g.][]{egu09}, to look for variations in the distribution of
different star formation tracers around the spiral arms. Standard
density-wave theory predicts a systematic trend in the distribution of
stellar ages inside and outside of the corotation
radius. \citet{foy11} used UV images from {\sl GALEX} and
IRAC/$3.6\mu$m emission to trace older populations. However, NGC\,4321
is not in their sample and the spatial resolution of UVOT is superior
to {\sl GALEX}.

In order to analyse the properties of the star forming sources with
respect to the spiral arms, we fit three portions of the spiral
structure of NGC\,4321 in regions where the arms can be
traced. Fig.~\ref{fig:spiral} shows the selection of the UV sources
along the spiral arms. We use the UVW2 and the H$\alpha$ images to
manually select a number of points along the spiral arms. The solid
lines tracing the arms in Fig.~\ref{fig:spiral} are obtained from a
linear interpolation of these sources in polar coordinates. The
crosses, triangles and circles locate the UVW2 sources that reside
within $\pm$2~kpc of the spiral arms.

\begin{figure}
  \includegraphics[width=8.8cm]{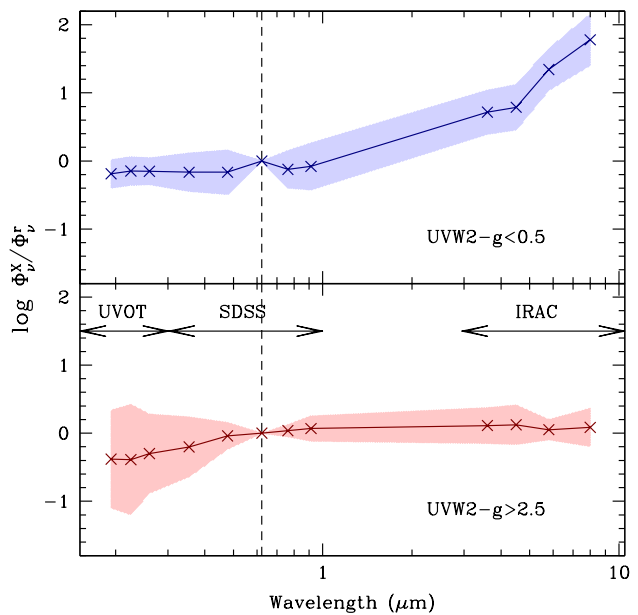}
  \caption{Spectral Energy Distribution of a subsample, selected
    according to UVW2-g colour as labelled. The lines and shaded areas
    map the average and standard deviation of all sources within the
    colour selection, shown as the flux through an ``X'' band
    representing the filters from UVOT to IRAC, as shown in the
    horizontal axis, measured as a ratio with respect to the SDSS r
    flux. The blue sources (top panel) have a clear excess towards
    longer wavelengths, possibly from PAH dust emission.}
  \label{fig:SEDs}
\end{figure}

Fig.~\ref{fig:DensWave} (left) shows a sketch of the changes expected
in the distribution of the spiral arm offsets of the star forming
sources with respect to age and galactocentric distance. The solid
blue and dashed red distributions illustrate the trend of the offsets
between a very young component (e.g. an HII region) and an older one
(e.g. an UV star forming source), respectively. The telltale sign of a
density wave with constant pattern speed would be the shift of the
older component with respect to the young one inside and outside
corotation. We emphasize here that we only focus on {\sl relative}
offsets between these two components, noting that there would be an
additional, smaller, offset between the HII regions and the molecular
gas, as found in, e.g. \citet{tamb08}. The histograms on the right
hand side of Fig.~\ref{fig:DensWave} show observed distributions of
various properties of the sources with respect to the distance to the
spiral arm for two choices of radial separation, at either side of the
corotation radius suggested by \citet{can97} (top panels: R$>$1.5h;
bottom panels: R$<$h).  They compared the two dimensional
observed velocity field with what the first-order linear spiral
density wave theory predicts, and measured a corotation radius R$_{\rm
  cor}=98\pm10$\,arcsec (shown for reference in
Fig.~\ref{fig:spiral}).  We note that this estimate, located around
R$\sim 1.5h$ falls in the middle of our distribution of sources. We
will adopt this value for reference in our analysis, although we note
that other studies give higher results: \citet{sem95} propose a
corotation radius in the range 8--11~kpc, corresponding to
100-150\,arcsec.  The pattern speed measured by \citet{rw04}, using
the Tremaine-Weinberg method, implies a corotation radius of $\sim
2$\,arcmin \citep{egu09}, while \citet{oey03} suggest a corotation
radius of $2.6$\,arcmin.  However, these methods may not be applicable
if the spiral arm pattern speed changes as a function of radius. For
instance, \citet{meidt09} applied their developed version of the
Tremaine-Weinberg method and demonstrated that some spiral galaxies
show a radial variation of the spiral pattern speed. Unfortunately,
NGC\,4321 is not in their sample.

\begin{figure}
 \includegraphics[width=8.5cm]{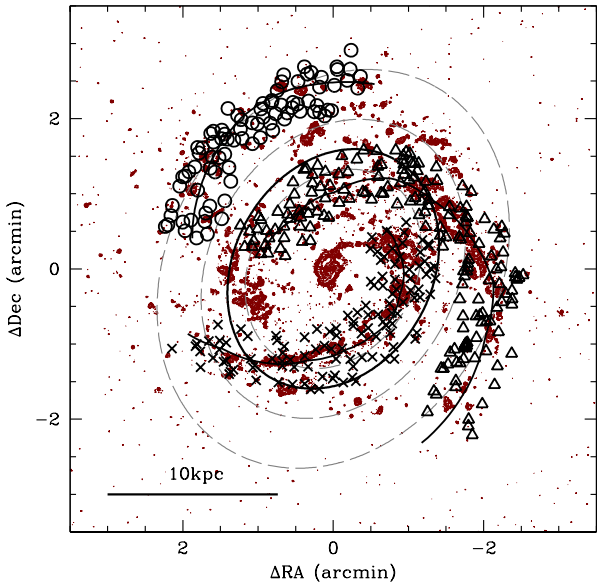}
 \caption{The H$\alpha$ sources are used to define three portions of
   the spiral arm structure where we can study the distribution of
   star forming sources. The contour map shows, as reference, the UVW2
   image, which tracks a wider range of stellar ages. The dashed
   projected circles have radii $h$, $3h/2$, and $2h$, where $h$ is
   the disc scale-length. The thick solid circle locates the
   corotation radius according to \citet{can97}.  The crosses,
   triangles and circles are the UVW2 sources located within
   $\pm$2~kpc of the definition of the arms.}
 \label{fig:spiral}
\end{figure}

The left and central panels are
model-independent comparisons of NUV-detected sources with respect to
their H$\alpha$ luminosity or UV--optical colour, respectively. The
H$\alpha$ selection is split at a reference value for the luminosity
of the line defined by $\log$\,L$_\alpha^0$(erg/s)$\equiv
37.9-0.4($UVW2$_{\rm AB}-20)$, motivated by the location of the
H$\alpha$ sources in the top panel of Fig.~\ref{fig:Rad}. Hence
L$_\alpha^0$ acts as a H$\alpha$-based tracer of stellar age. The
middle panel is motivated by the broadband photometric analysis (see
Figs.~\ref{fig:Ages} and \ref{fig:SEDs}) but does not rely on any
modelling.  Finally, the rightmost panels separate the sample
according to the ages of the young components according to this
analysis, and described in the previous section.  No significant
offsets are found in any of these cases.

\begin{figure*}
  \includegraphics[width=8cm]{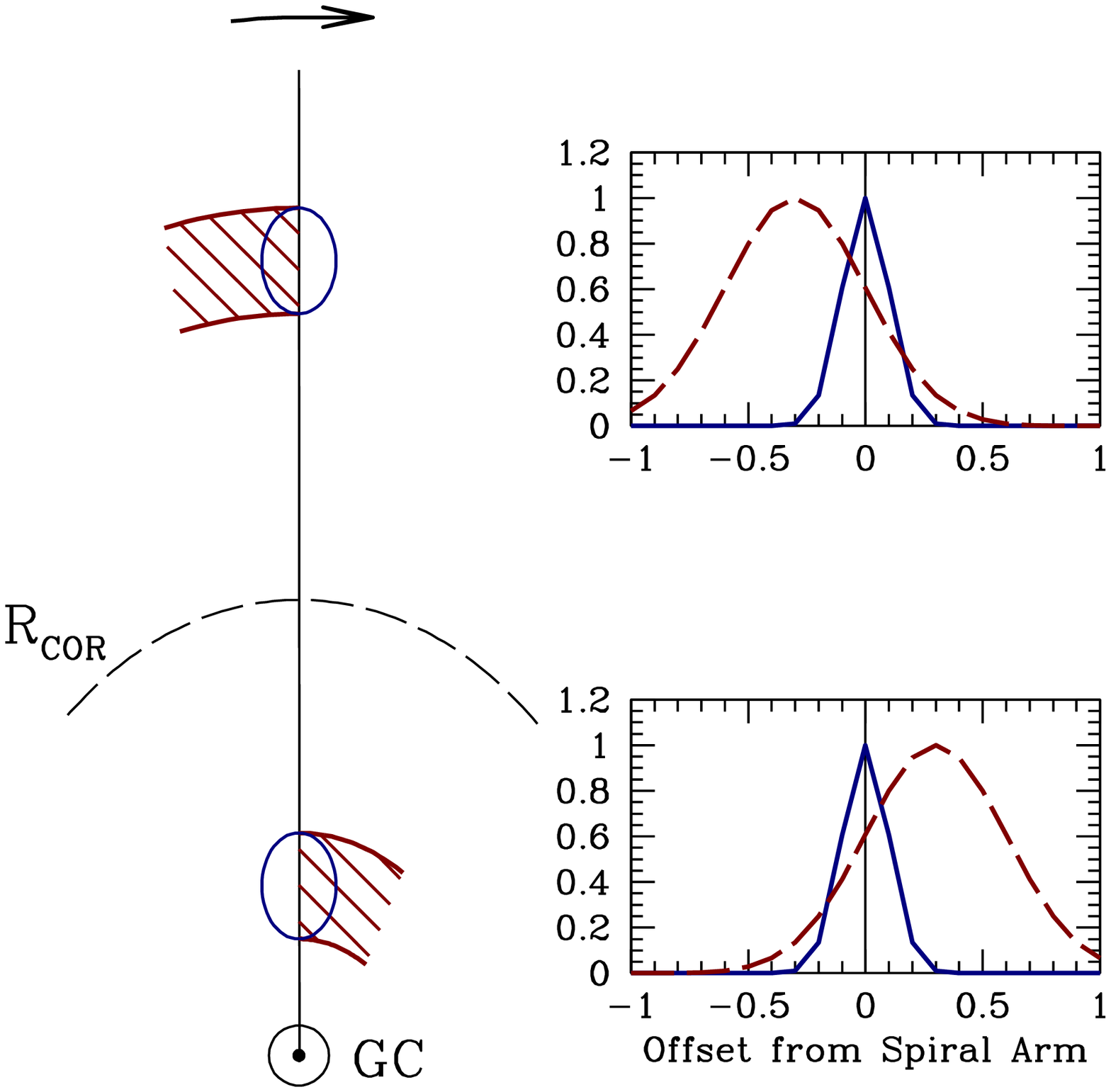}
  \includegraphics[width=8cm]{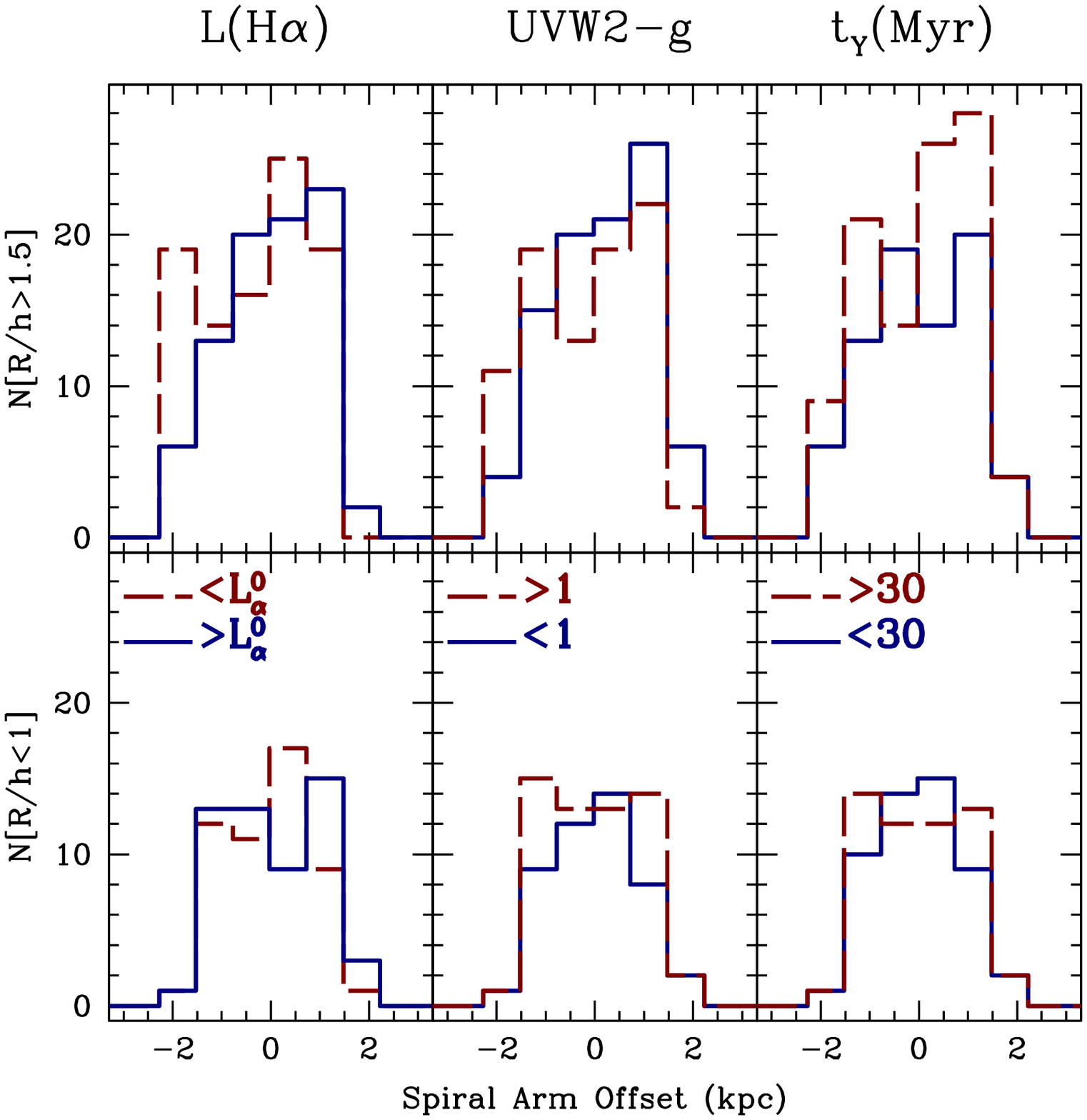}
  \caption{{\sl Left:} Sketch of the expected shifts in the
    distribution of stellar populations according to density wave
    theory. A spiral pattern with constant angular speed --
    represented by the vertical line -- interacts in different ways
    with the rotating gas depending on whether the gas is located
    inside or outside the corotation radius R$_{\rm COR}$. On the
    left panels, a comparison of very recent star formation -- as
    tracked by H$\alpha$ (solid blue curves) -- with older stars --
    from UV light (dashed red curves) betrays the presence of a
    pattern in solid body rotation.  {\sl Right:} Comparison between
    the distribution of UVW2-selected sources split according to
    H$\alpha$ line luminosity (L(H$\alpha$), leftmost panels),
    UV-optical colour (UVW2--g, middle panels) or the age of the young
    component (t$_{\rm Y}$, rightmost panels), estimated from the
    modelling of the NUV/Optical/IR photometry. The threshold in
    H$\alpha$ luminosity, L$_{\alpha}^0$, is defined with respect to the UVW2
    flux in Fig.~\ref{fig:Rad}.  Only sources within $\pm$2\,kpc from
    the three spiral arms defined in figure~\ref{fig:spiral} are
    included in the histograms, split according to radial distance to
    the centre, measured in disc scale lengths.}
  \label{fig:DensWave}
\end{figure*}

The lack of an offset is clearly inconsistent with a solidly rotating
spiral arm pattern as predicted by density wave theory. \citet{dp10}
show a predicted distribution of young star clusters with different
ages from their numerical simulations with a fixed spiral arm
potential rotating solidly and mimicking a long-lived spiral
arm. Their fig.~4 provides a quantitative plot of the sketches we
outlined on the left of our Fig.~\ref{fig:DensWave} and they consider
alternative mechanisms for the formation of spiral arms that would
give rise to similar distributions as our observed
histograms. Interestingly, \citet{egu09} also found no offset between
CO and H$\alpha$ spiral arms in NGC\,4321, although we note that their
analysis covered a narrower radial range (R$<1.4^\prime$, roughly one
scale length). In combination with our analysis, we can therefore
state that in NGC\,4321 no offset is found out to 2 scale lengths over
a wide range of ages, from molecular clouds to about 100 Myr young
stars. This is a striking result and agrees with what is also found in
\citet{foy11} for other galaxies. Our result provides further evidence
against classic spiral density wave theory.

\begin{table*}
\caption{Grid of 2-population models used in the analysis}
\label{tab:2bst}
 \begin{tabular}{lccc}
\hline
Parameter & Young SSP & Old SSP & Models\\
\hline
Age (Gyr) & $-3<\log t_{\rm Y}<0$ & $0<\log t_O<+1.12$ & $64\times 64$\\
Metallicity (Z$_\odot$) & \multicolumn{2}{c}{$-2<\log Z<+0.3$} & 24\\
Dust Reddening & $0<E_{B-V}<1$& dustless & 32\\
Mass fraction & $0<f_{\rm Y}<0.8$ & $1-f_{\rm Y}$ & 32\\
\hline
\multicolumn{3}{r}{TOTAL} & $1.007\times 10^8$\\
\end{tabular}
\end{table*}


The distributions found in Fig.~\ref{fig:DensWave} can be naturally
explained if the spiral arm is ``corotating'' with the stars and the
gas, i.e. if the spiral pattern speed decreases with galactocentric
radius. Recent numerical simulation studies \citep{wada11,rob12} claim
that such corotating spiral arms are more naturally developed in
numerical simulations. Indeed, it is rather difficult to reproduce the
spiral density wave theory in numerical simulations with self-gravity
\citep{sellwood11}. Such corotating spiral arms suffer from the winding
problem, but are found to be disrupted within a time-scale of a few
hundreds of Myr \citep[see e.g.][and references
  therein]{fuj11,quill11,sellwood11} and a snapshot of the evolving
galaxy at any time always shows some spiral arm structure.
\citet{wada11} also demonstrate that in such corotating spiral arms
gas falls into the spiral arms both from behind and the front of the
spiral arm and collides, inducing star formation in the arm. This
scenario is consistent with the observed distribution of star forming
sources presented here.

\section{Conclusions}

The resolution of the {\sl Swift}/UVOT camera makes it an optimal
instrument to explore the young star forming sources in nearby
late-type galaxies.  In combination with optical photometry, it is
possible to disentangle the contribution from the more diffuse, older
component, giving an estimate of the age and dust content of the young
stars responsible for the NUV emission.  In this study, we combine
deep UVOT imaging in the NUV with optical and IR archival imaging to
study the properties of 787 star forming regions in the spiral
structure of NGC\,4321, out to R$\sim$3h. We find a strong trend of
brighter H$\alpha$ sources in the central regions
\citep[Fig.~\ref{fig:Rad}, see also][]{godd11}, whereas the ages, or
the dust content of the star forming regions do not present a
measurable correlation with galactocentric distance (Fig.~\ref{fig:Ages}),
suggesting that the strong trend in H$\alpha$ flux is caused by a
larger number of massive star forming sources in the central regions
(assuming a universal stellar initial mass function).  
The presence of dust is revealed when stacking up the photometric SEDs
of the sources according to their NUV-optical colour, with significant
excess at $8\mu$m for the bluest (and youngest) ones
(Fig.~\ref{fig:SEDs}). However, a quantitative analysis of the dust
reddening from the photometric data gives rather low ``spatial resolution
averaged'' values of E(B--V)$\simlt$0.2~mag.
The main outcome of the comparison with stellar populations is that
one can use UVW2--g colours as a proxy of the age of the younger
component, ranging from a few Myr to a few hundred Myr (see
Fig.~\ref{fig:Ages}).

The properties of these sources enable us to study the effect of the
spiral structure, a powerful test of pattern formation in disc
galaxies. The standard scenario of density wave theory with a constant
pattern speed results in an offset with respect to age for the
distribution of distances to the spiral arms as one moves from the
central regions -- where the gas moves faster than the pattern -- to
the outer disc (see Fig.~\ref{fig:DensWave}, left).  The advantage of
UV imaging over studies based on comparisons between the gas component
and HII/$24\mu$m regions is the wider time baseline spanned by UV
light, roughly of order 100~Myr. No significant differences are found
in the distribution of these sources, giving further negative evidence
for density wave spirals. Our findings reject the possibility that the
results from \citet{egu09} -- giving no offsets between CO and
H$\alpha$ over a narrower range (R$<1.4^\prime$) -- can be
explained by a radial coincidence with corotation. Instead, a scenario
with short-lived corotating spiral arms is compatible with the
data. Such structures have been found in numerical simulations of
spiral galaxies \citep[see e.g.][]{rob12}, implying a fundamental
change of our views from the traditional concept of long-lived spiral
arms.

\section*{Acknowledgments}
We would like to thank the MSSL {\sl Swift}/UVOT team for useful help
on the data reduction of the UV datasets. The constructive criticism
of the anonymous referee has greatly helped in improving this paper.

\noindent
Funding for SDSS-III has been provided by the Alfred P. Sloan
Foundation, the Participating Institutions, the National Science
Foundation, and the U.S. Department of Energy Office of Science. The
SDSS-III web site is http://www.sdss3.org/.  SDSS-III is managed by
the Astrophysical Research Consortium for the Participating
Institutions of the SDSS-III Collaboration including the University of
Arizona, the Brazilian Participation Group, Brookhaven National
Laboratory, University of Cambridge, University of Florida, the French
Participation Group, the German Participation Group, the Instituto de
Astrofisica de Canarias, the Michigan State/Notre Dame/JINA
Participation Group, Johns Hopkins University, Lawrence Berkeley
National Laboratory, Max Planck Institute for Astrophysics, New Mexico
State University, New York University, Ohio State University,
Pennsylvania State University, University of Portsmouth, Princeton
University, the Spanish Participation Group, University of Tokyo,
University of Utah, Vanderbilt University, University of Virginia,
University of Washington, and Yale University.


\label{lastpage}
\end{document}